# A Framework for Optimizing Paper Matching

**Laurent Charlin**
Department of Computer Science
University of Toronto
Toronto, ON M5S 3H5
*lcharlin@cs.toronto.edu*

**Richard Zemel**
Department of Computer Science
University of Toronto
Toronto, ON M5S 3H5
*zemel@cs.toronto.edu*

**Craig Boutilier**
Department of Computer Science
University of Toronto
Toronto, ON M5S 3H5
*cebly@cs.toronto.edu*

## Abstract

At the heart of many scientific conferences is the problem of matching submitted papers to suitable reviewers. Arriving at a good assignment is a major and important challenge for any conference organizer. In this paper we propose a framework to optimize paper-to-reviewer assignments. Our framework uses suitability scores to measure pairwise affinity between papers and reviewers. We show how learning can be used to infer suitability scores from a small set of provided scores, thereby reducing the burden on reviewers and organizers. We frame the assignment problem as an integer program and propose several variations for the paper-to-reviewer matching domain. We also explore how learning and matching interact. Experiments on two conference data sets examine the performance of several learning methods as well as the effectiveness of the matching formulations.

## 1 Introduction

The assignment of papers to reviewers is one of the most important tasks facing the organizers of scientific conferences. Assigning submitted papers to their most suitable reviewers is essential to the success of any conference, indeed to the functioning of many scientific fields, since it is reviewer assessments that determine the conference program and, to some extent, the shape of a discipline. However, this is not a simple task: large conferences often receive well over 1000 submissions that must be assigned to many hundreds of reviewers in short amount of time. Apart from ensuring the suitability of assigned reviewers, constraints imposed by reviewer load limits, conflicts of interest, and other factors push this assignment problem beyond the reach of a single program chair, and generally prevent the process from being distributed in a fully satisfactory way.

Many large conferences in computer science (CS), and especially artificial intelligence (AI), allow reviewers to bid on papers, basically providing their "preferences"—which we interpret as reflecting a reviewer's *suitability* to review particular submitted papers—after which a centralized matching process takes place to find the most suitable assignment. Preferences collected this way are, unfortunately, inherently noisy for two key reasons: (a) it is difficult for reviewers to offer reasonable assessments of all but a small fraction of the papers, given the numbers involved; and (b) reviewers have access to limited information about each paper (e.g., only title and abstract). The latter factor fundamentally limits how well a reviewer can judge her own suitability, while the former means reviewers are, in some sense, semi-randomly choosing papers on which they express interest.

One response to this problem is to associate simple features (keywords being most common) with both papers and reviewers, and use some measure of overlap as a reflection of suitability. Unfortunately, this simple method is crude at best, and relies on a common understanding of this (usually limited) vocabulary by *all reviewers and authors*. A more sophisticated response involves the use of machine learning techniques to help predict reviewer expertise [1, 16, 6]. By using specific features of both reviewers (e.g., previously written articles, co-authorship relations) and submitted papers (e.g., words or keywords), we can relieve reviewers of the burden of bidding. Ideally, a combination using this information as well as self-declared reviewer expertise (or bids) can be leveraged to predict reviewer suitability using collaborative filtering methods [10]. Ultimately, however, the primary goal is not to accurately predict expertise, but to find a good matching.

In this paper, we propose and test various instantiations of a flexible framework for optimizing paper matching. We investigate approaches that use incomplete information in the form of a limited number of suitability scores: our basic framework predicts missing scores using learning techniques and then finds optimal matchings using both observed and predicted scores. Within this framework, we explore several learning models which leverage (one or both of) two sources of information—reviewer/paper features and self-reported suitability—to predict the unknown scores: these include regression, collaborative filtering and language modeling methods. We then describe several desirable properties for paper-to-reviewer assignments. We frame the assignment problem as an integer program [24], and explore several variations that reflect different desiderata, and how these interact with various learning methods. We test our framework on two data sets collected from a large AI conference, measuring predictive accuracy with respect to both reviewer suitability and matching performance, exploring several different matching objectives and how they can be traded off against one another.

Although we focus on reviewer matching, our methods are applicable to any constrained matching domain where: (a) user preferences for a set of items can be predicted using user and/or item features; (b) preferences can be used to improve matching quality; (c) it is infeasible or undesirable for users to express preferences over all items; and (d) capacity or other constraints limit the min/max number of users-per-item (or vice versa). Examples include facility location, school/college admissions, certain forms of scheduling and time-tabling, and many others.

## 2 Related Work

Deep bodies of related work exist for each of the two components that comprise our framework for reviewer matching: prediction of suitabilities or preferences for unobserved reviewer-paper (or user-item) pairs; and computing matchings given (known or predicted) suitabilities. Past work has either explored the score prediction problem or different approaches to matching but, to the best of our knowledge, ours is the first that examines suitability prediction relative to different matching objectives, and examines the interactions between learning and matching.

There has been significant research on the use of information retrieval and learning techniques to determine suitability of reviewers for papers. These include the use of latent semantic indexing [7] or term frequency, inverse document frequency (TF-IDF) methods [12, 2] that exploit the content of abstracts of papers authored by reviewers and those of submitted papers. Other have utilized co-authorship graphs, using the references of a submitted paper as a starting point to generate potential referees [18]. Balog et al. [1] used language models to determine the suitability of experts for various topics/tasks, and more recently topic models have been applied to the problem of modeling expertise based on authored documents [25], with Mimno and McCallum [16] applying their topic model to the assessment of reviewer suitability (we discuss this further below).

While the models above predict suitability using content-based features of papers and/or reviewers, other methods exploit elicited suitability scores from reviewers for a subset of papers to make predictions for other papers. This can be treated as a *collaborative filtering (CF)* problem. CF methods leverage known preference information for a subset of user-item to generate predictions for unobserved pairs. Recent CF techniques have performed extremely well in a variety of domains, especially where available content features are not especially predictive of preference (or suitability) [21, 22, 15]. Conry et al. [6] applied an ensemble CF approach, combining side information about the papers and reviewers with several CF predictors to estimate reviewer suitabilities, and then used a simple matching program to determine assignments based on these suitabilities. This work is closest to ours; however, it does not explore variants of the matching objective, nor interactions between learning and matching. While CF is typically framed in terms of preference prediction, recent extensions instead use CF for optimization w.r.t. a specific target task. Weimer et al. [26] use CF data for optimization in a ranking task, while Petterson et al. [17] frame ranking as finding the weights that lead to an optimal matching in a bipartite graph. Our work has a similar motivation, trying to optimize suitability predictions w.r.t. a matching objective.

A second body of work focuses on the matching problem itself. Benferhat and Lang [3], Goldsmith and Sloan [11], and Garg et al. [9] discuss various optimization criteria, and some of the practices used by program chairs and existing conference management software. Taylor [24] shows how these criteria can be formulated as an integer program (IP). Tang et al. [23] propose several extensions to the IP. This work assumes reviewer suitability for each paper is known, and deals exclusively with specific matching criteria. There is a rich literature on more general matching problems in economics and theoretical CS. Examples include the well-known *stable marriage problem* [8]; resident matching (of residency candidates to hospitals) [19]; and (one-sided) matching in housing markets [13]. In economic models, a key focus is on stability of the matches and minimizing incentives for participants in the matching market to misreport their preferences. We do not consider such strategic issues here.

## 3 Matching Framework and Instantiations

We begin by outlining our basic problem definition, then elaborate on several specific instantiations of the framework we develop. These include the use of various learning methods for predicting unknown suitabilities, a range of objectives and constraints on the matching process reflecting different desiderata for the reviewing process, and interactions between the two.

### 3.1 Problem Definition

Our approach to the matching problem relies on *suitability scores*, which describe the relevance of a reviewer to a given paper. The matching procedure uses these scores to form a set of assignments of items to users. For reasons discussed above, the suitabilities will not be fully specified. Since we do not wish to limit the matching process to reviewer-paper pairs that are known (i.e., have been directly elicited), these need to be predicted in some fashion.

We formalize the matching problem as follows. Let $r \in \mathcal{R}$ refer to users or *reviewers*, $p \in \mathcal{P}$ to items or *papers*, and let $|\mathcal{R}| = N$ and $|\mathcal{P}| = M$. Every user-item pair has a *suitability score* $s_{rp}$. The set of all scores can be viewed as a suitability matrix $S \in \mathbb{R}^{N \times M}$. Only a subset of the suitabilities are observed, namely, those collected from reviewers during an elicitation process. Denote this by $S^o$, and denote the observed scores for a particular reviewer $r$ and paper $p$ by $S^o_r$ and $S^o_p$, respectively. $S^u, S^u_r, S^u_p$ are the analogous collections of unobserved scores.

We may have access to additional side information about individual reviewers and papers which may come in different forms. In our setting, side information about submitted papers could include author-specified keywords, citations, and word usage in the paper. For reviewers, we may have stated preferences for keywords, citations, or other descriptions of reviewer expertise. Our data sets also include an *archive*, containing a set of papers written by each reviewer, providing information about their expertise. This is represented as a word count vector $w_r$ summarizing $r$'s own papers. Similarly, we summarize each submitted paper $p$ as a word count vector $w_p$.

Given this information, our goal is to find a "good" matching of papers to reviewers in the presence of incomplete information about reviewer suitabilities, possibly exploiting the side information available. The problem can be broken into two main components: predicting unknown suitabilities using some combination of known scores and side information; and matching papers to reviewers based on known and predicted suitabilities. Notice that predicting suitability scores is, however, not a goal in and of itself: it is subservient to the primary goal of good matching performance. Many different factors may be used to define the quality of a matching, as we discuss below.

### 3.2 Learning Methods

We have explored a range of learning methods for predicting suitabilities of reviewers for papers. Here we focus on three methods, each exploiting the different information available for prediction: a *language model (LM)*; *linear regression (LR)*; and *Bayesian probabilistic matrix factorization (BPMF)*. LM uses the content of submitted papers and archived papers for prediction, but does not use reviewer bids; BPMF uses reported suitabilities/bids, but no document/archive side information; and LR uses bids and the content submissions, but not the archive.

**Language Model:** Several previous approaches to reviewer matching have used simple language models to represent distributions over words of papers and reviewers [25, 16]. Our language model (LM) is based on these, and predicts suitabilities without using stated reviewer preferences; rather it builds a model in word (feature) space, assuming that distance in this space correlates with distance in suitability space. LM constructs a distribution over words for each reviewer, based on the archive of papers written by the reviewer $w_r$ (the reviewer side information). The starting point for LM is a multinomial $\Pr(w|d)$ over words $w$ in a document $d$. The maximum likelihood estimate of $\Pr(w|d)$ is the number of occurrences of this word divided by the total number of words in the document ($\Pr_{ml}(w|d) = |w_d|/T_d$). Using Dirichlet smoothing to account for the fact that most words do not appear in a given document, this estimate can be written as:

$$\Pr(w|d) = \frac{T_d}{T_d + \mu}\Pr_{ml}(w|d) + \frac{\mu}{T_d+\mu} \Pr(w) \quad (1)$$

where $\Pr(w)$ is the probability of the word across all documents and $\mu$ is the smoothing parameter. This distribution can be formed in various ways from the user side information (i.e., the collected papers of a reviewer). We adopt a variant of an approach [16] in which the word vectors of reviewer-authored papers are averaged to form a single document $d_r$ per reviewer. LM encodes each submitted paper as a word count vector $w_p$, and predicts suitabilities $s_{rp}$ to be $\log \Pr(w_p|d_r) = \sum_{w \in w_p} \Pr(w|d_r)$. This language model has outperformed sophisticated topic models in some settings [16].

**Regression:** Linear regression (LR) predicts suitabilities directly using the side information associated with the items. Each reviewer has a set of parameters $\theta_r$, which is applied to item information $w_p$ to form an estimate of $s_{rp}$: $\hat{s}_{rp} = \theta_r \cdot w_p$. Stated reviewer preferences are used as training observations, and LR minimizes the mean-squared error (MSE) w.r.t. observed suitabilities:

$$C_{LR}(S^o) = \frac{1}{|S^o|} \sum_{s_{rp} \in S^o} (\hat{s}_{rp} - s_{rp})^2 \quad (2)$$

**Collaborative Filtering:** Given observed suitabilities, prediction of unobserved suitabilities can be tackled using collaborative filtering. *Probabilistic matrix factorization (PMF)* [21] is a popular CF method that finds a low-rank factorization of the suitability matrix $S \approx U^T V$, where $S \in \mathbb{R}^{N \times M}$, $U \in \mathbb{R}^{K \times N}$ and $V \in \mathbb{R}^{K \times M}$ and $K << \min(M, N)$. The columns $U_r$ of $U$ and $V_p$ of $V$ represent latent reviewer and paper factors. The full $S$ matrix, including unobserved suitabilities, can be estimated by taking the product of $U$ and $V$. Under this model, the conditional distribution over suitabilities is:

$$\Pr(S|U,V,\sigma^2) = \prod_r^M \prod_p^N \mathcal{N}(s_{rp}|U_r^T V_p, \sigma^2)^{I_{rp}}$$

where $I$ is an indicator matrix and entry $I_{rp}$ is 1 if it was observed and 0 otherwise. Assuming zero-mean Gaussian priors over the parameters $U$ and $V$:

$$\Pr(U|\sigma_U^2) = \prod_{r=1}^N \mathcal{N}(U_r|0,\sigma_U^2); \quad \Pr(V|\sigma_V^2) = \prod_{p=1}^M \mathcal{N}(V_p|0,\sigma_V^2),$$

finding the MAP solution involves minimizing MSE between $U^T V$ and the known suitabilities.

In a Bayesian version of PMF (BPMF), the parameters $U$ and $V$ have non-zero mean and full-covariance priors [22]. The predictive distribution cannot be calculated analytically because the posteriors over $U$ and $V$ are intractable, but a Markov Chain Monte Carlo sampler can be used to approximate sufficient statistics. Integrating over the parameters has been shown to produce performance advantages w.r.t. root mean squared error (RMSE) on the Netflix task [22].

**Other Methods:** In addition to the three methods above, we investigated several other algorithms, including supervised and unsupervised topic models [4, 5], conditional restricted Boltzmann machines (RBMs) [14], and inference using co-reference graphs. Preliminary experiments showed that these methods did not match the performance of those above. We also tried leveraging the different sources of information by using combinations of the various learning models without success. Incorporating each reviewer's archive as extra training papers for LR did not offer any improvement either. Finally, we explored a formulation of the problem in which the training objective explicitly optimizes matchings $J(x)$ rather than optimizing RMSE w.r.t. predicted suitabilities. However, experiments with this approach failed to demonstrate improved matching performance.

### 3.3 Matching Objectives

We articulate several different criteria that may influence the definition of a "good" matching and explore different formulations of the optimization problem that can be used to accommodate these criteria. We also discuss how these criteria may interact with our learning methods.

Naturally, one would like to assign submitted papers to their most suitable reviewers; of course, this is almost never possible since some reviewers will be most suited to far more papers than others. In general, *load balancing* is enforced by placing an upper limit or *maximum* on the number of papers per reviewer. Similarly, we may impose a *minimum* to ensure reasonable *load equity* or load fairness across reviewers. However, limiting the paper load increases the probability that certain papers will be assigned to very unsuitable reviewers. This suggests only making assignments involving pairs with score $s_{rp}$ above some *minimum score threshold*. This ensures that every paper is reviewed by a minimally suitable reviewer, but may sacrifice load equity (indeed, it may sacrifice feasibility). One may also desire *suitability fairness* across reviewers; that is, reviewers should have similar score distributions over their assigned papers (so on average no reviewer is assigned papers to which she is significantly more ill-suited than any other). Finally, when multiple reviewers are assigned to papers, it may be desirable to assign *complementary reviewers* to a paper so as to cover the range of topics spanned by a submission. Related is the desire to ensure each paper is reviewed by *at least one "senior" reviewer* with significant expertise.

The intricacies of different conferences prevent us from establishing an exhaustive list of matching desiderata (see [3, 11, 9] for further discussion). We now explore matching mechanisms that will account for several of these criteria: we frame the matching procedure as an optimization problem and show how several properties can be formulated as constraints or modifications of the objective function.

We formulate the basic matching problem as an IP, where each paper is assigned to its best-suited reviewers [24]:

$$\text{maximize} \quad J^{basic}(x) = \sum_r \sum_p s_{rp} x_{rp} \quad (3)$$

$$\text{subject to} \quad x_{rp} \in \{0, 1\}, \quad \forall r, p \quad (4)$$

$$\sum_r x_{rp} = R_{target}, \quad \forall p$$

The binary variable $x_{rp}$ encodes the matching of item $p$ to user $r$; a *match* is an instantiation of these variables. $R_{target}$ is the desired number of reviewers per paper. Minimum and maximum reviewer load, $P_{\min}$ and $P_{\max}$ respectively, can be incorporated as constraints [24]:

$$\sum_p x_{rp} \geq P_{\min}, \quad \sum_p x_{rp} \leq P_{\max}, \forall r. \quad (5)$$

This IP, including constraints (5), is our basic formulation (*Basic IP*). Its solution, the *optimal match*, maximizes total reviewer suitability given the constraints. Although IPs can be computationally difficult, our constraint matrix is totally unimodular, so the linear program (LP) relaxation (allowing $x_{rp} \in [0, 1]$) does not affect the integrality of the optimal solution; hence the problem can be solved as an LP.

Although not mentioned above it is essential for the matching to prevent assignments of reviewers to submitted papers for which they have conflicts of interest (COI). The above formulation can easily enforce known COI by directly constraining the conflicting assignments $x_{rp}$'s to be 0, alternatively, we can set the relevant scores $s_{rp}$'s to $-\infty$.

To capture additional matching desirata, we can modify the objective or the constraints of this IP. Load balancing can be controlled by manipulating $P_{\min}$ and $P_{\max}$: a small range ensures each reviewer is assigned the same number of papers at the expense of match quality, while a larger range does the converse. We can instead enforce load equity by making the tradeoff explicit in the objective with "soft constraints" on load:

$$J^{balance}(x) = \sum_r \sum_p s_{rp} x_{rp} + \sum_r \lambda f\Big(\Big(\sum_p x_{rp}\Big) - \bar{x}\Big) \quad (6)$$

where $\bar{x}$ is the average number of papers per reviewer ($M/N$) and $f$ is a penalty function (e.g., $f(x) = |x|$ or $f(x) = x^2$). The parameter $\lambda$ controls the tradeoff between load equity and match quality. The $J^{balance}$ objective (Eq. 6) along with the constraints expressed in Eq. 4 comprise our *Balance IP*.

The $J^{basic}$ objective (Eq. 3) maximizes the overall suitability of the assignments, equating "utility" with suitability. However, the utility of a specific match $x_{rp}$ may not be linear in suitability $s_{rp}$. For example, utility may be more "binary": as long as a paper is assigned to a reviewer whose suitability is above a certain threshold, then the assignment is good, otherwise it is not. This can be realized by applying some non-linear transformation $g$ to the scores in the matching objective (e.g., a matched pair with score $s_{rp} \in \{2, 3\}$ may be greatly preferred to $s_{rp} \in \{0, 1\}$):

$$J^{tfm}(x) = \sum_r \sum_p g(s_{rp}) x_{rp}. \qquad (7)$$

In this *transformed* objective $J^{tfm}$, if $g$ is a logistic function then score are softly "binarized."

Finally, we note that some of these matching objectives can also be incorporated into the suitability prediction model. For example, the nonlinear transformation $g$ can be directly used in the LR training objective (cf. Eq. 2):

$$C_{\text{LR-TFM}}(S^o) = \frac{1}{|S^o|} \sum_{s_{rp} \in S^o} (\hat{s}_{rp} - g(s_{rp}))^2. \qquad (8)$$

## 4 Experimental Results

We start by describing the data sets used in our experiments. The rest of the section is divided into three parts. The first considers score predictions with the different learning models. The second turns to matching quality and explores the soft constraints on the number of papers matched per reviewer. Finally, the third part evaluates a transformation of the matching objective and shows how using a transformed learning objective can enhance performance on the transformed matching problem.

### 4.1 Data

Experiments are run using two data sets, *N10* and *N09*, from the 2010 and 2009 editions, respectively, of the NIPS conference, one of the leading conferences in machine learning.[1] For both data sets, side information for each reviewer comprises a self-selected set of papers taken representative of her areas of expertise; these were summarized as word count vectors $w_r$. Side information about submitted papers consisted of document word counts $w_p$ for each $p$. The total vocabulary used by submissions (across both sets) contained over 21,000 words; We used only the top 1000 words for our experiments as ranked using TF-IDF ($|w_p| = |w_r| = 1000$). Reviewer suitability scores ranged from 0 to 3, with 0 meaning "paper lies outside my expertise;" 1 means "can review if necessary;" 2 means "qualified to review;" and 3 means "very qualified to review." As discussed above, these scores are intended to reflect reviewer expertise, not desire. We focus on the area chair (or meta-reviewer) assignment problem, where the matching task is to assign a single area chair to each paper. We use the term reviewer below to refer to such area chairs.

[1] See http://nips.cc. We are currently investigating mechanisms by which we can make both data sets available to the community.

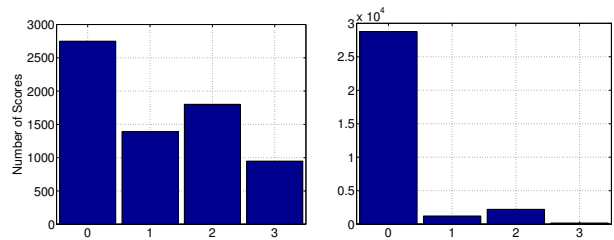

Figure 1: Observed scores for *N10* (a) and *N09* (b).

*N10* comprises 1250 submitted papers to be assigned to 48 reviewers. Suitabilities on a subset of papers were elicited from reviewers using a rather involved two-stage process. This process utilized the language model (LM) to estimate the suitability of each reviewer for each paper, and then queried each reviewer on the papers on which his estimated suitability was maximal. The output of the first round was fine-tuned using a combination of a hybrid discriminative/generative RBM [14] with replicated softmax input units [20] trained on the initial scores, and LM, which then determined the second round of queries. In total, each reviewer provided score on an average of 143 queried papers (excluding one extreme outlier), and each paper received an average of 3.3 suitability assessments (with a std. dev. of 1.3). The mean suitability score was 1.1376 (std. dev. 1.1); a histogram of the scores is shown in Figure 1(a). Note that since the querying process was biased towards asking about pairs with high predicted suitability, the *unobserved* scores are not missing at random, but rather tended toward pairs with *low* suitability. We do not distinguish the data acquired in the two phases of elicitation; both took place within a short time frame, so we assume suitabilities for any one reviewer are stable.

*N09* comprises 1079 submitted papers and 30 area chairs. Reviewer scores were not elicited, but instead provided by the conference program chairs for every reviewer-submission pair. A histogram of the scores is shown in Figure 1(b). The mean suitability score was 0.19 (std. dev. 0.57).

### 4.2 Score Predictions

We first analyze performance using the root mean squared error (RMSE) metric, as is common in collaborative filtering research. We are especially interested in how the different approaches behave as the size of the training set increases. An understanding of this dynamic is vital if one is to strike a balance between the demands of eliciting suitability scores from the user and increased accuracy of predictions.

For learning, we are given a set of training instances, $S^{tr} \equiv S^o$. We split this set into a training and validation set. The trained model predicts all unobserved scores $S^u$.

Since we do not have true suitability values for all unobserved scores, we distinguish $S^u$ as being the union of test instances $S^{te}$ (for which we have scores in the data set), and missing instances $S^m$. We denote a model's estimates of the test instances as $\hat{S}^{te}$, and evaluate RMSE over these test instances $(\sum_{rp \in S^{te}} (\hat{s}_{rp}^{te} - s_{rp}^{te})/|S^{te}|)^{1/2}$.

We report results averaged over 5 different splits of the data in all experiments. In each split, the data is divided into training, validation and test sets in 60/20/20 proportions. There is no overlap in the test sets across the 5 splits. When reporting results across splits, we report the mean and standard error. Training LR is naturally slightly faster than training BPMF [2], for which we used 330 MCMC samples including 30 burn-in samples, but both methods are trained within minutes on both of our data sets.

Fig. 2 shows results for training sets of different sizes, simulating the effect of additional elicitation. To facilitate comparison, the test set size is fixed across different training set sizes. We compare LR and BPMF to a baseline which predicts the mean training score. Recall that BPMF learns to predict $S^u$ by using $S^o$ only while LR also utilizes paper features $w_p$. The strong performance of LR suggests that the information contained in the paper features is extremely useful in predicting user preferences. Interestingly, BPMF, a state-of-the-art method in CF, performs worse than LR in all but the extremely small training-set sizes. Recall that BPMF attempts to exploit similarities across reviewers and across papers. In this case, the (meta-)reviewers were specifically chosen by the conference organizers to span the field, providing expertise across the multitude of topics typically represented at this conference. We conjecture that, compared to other popular domains for CF (e.g., movie recommendation), there are far fewer commonalities across users (w.r.t. paper topics); thus it is difficult for BPMF to attain very good performance. Furthermore, although there are probably significant commonalities between papers, each paper receives an average of fewer than four ratings, which makes it difficult to discover those commonalities from the preference data alone.

Note that the language model is not included here. LM's outputs values represent log-probabilities and thus do not fall into the $[0, 3]$ score range. A linear mapping of the predicted suitabilities into the score range did not produce sensible results.

The behavior of the different learning methods on *N09* are similar to that observed for *N10*. Since they provide no additional insight, we do not report results here.

### 4.3 Match Quality

We now turn our attention to the matching framework. We first elaborate on how we perform the matching. We then evaluate the performance of the different learning methods on the matching objective. Finally we introduce soft constraints into the matching objective and analyze the tradeoffs they introduce.

---

[2] We use an implementation of BPMF provided by its authors: http://www.mit.edu/~rsalakhu/BPMF.html

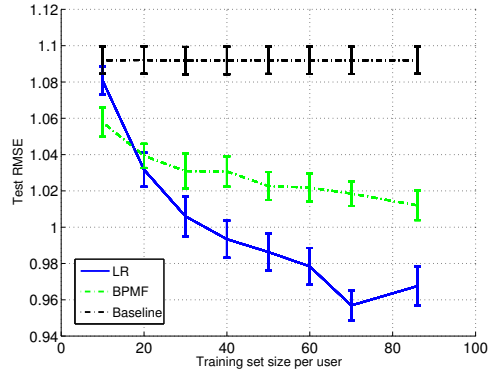

Figure 2: Performance on RMSE on the *N10* data set as the number of observed suitability scores per user varies.

|  | train/valid. | test | missing |
|---|---|---|---|
| Matching | $S^{tr}$ | $\hat{S}^{te}$ | $S^m = \tau$ |
| Evaluation | $S^{tr}$ | $S^{te}$ | $S^m = \tau$ |

Table 1: Overview of the matching/evaluation process.

**Matching Experimental Procedures:** The matching IPs discussed above assume access to fully known (or predicted) suitability scores. Since we learn estimates of the unknown scores, we denote a model's estimates of the test instances as $\hat{S}^{te}$, and impute a value for all suitability values that are missing, using a constant imputation of $\tau \in \mathbb{R}$. Since missing scores are likely to reflect, on average, lower suitability than their observed counterparts, we use $\tau = 1$ in all experiments (recall that *N10*'s mean was 1.1376 and *N09* has no missing scores).

Given the estimate $\hat{S}^{te}$ computed by one of our learning methods, we perform a matching with $S = S^{tr} \cup \hat{S}^{te} \cup (S^m = \tau)$. Note that this permits missing values to be matched, which is important in the regime of sparse known-suitability scores. Table 1 summarizes this procedure. For data set *N10* we set $P_{\min}$ and $P_{\max}$ to 20 and 30, respectively, while the range is 30–40 for data set *N09*.

*Baseline:* We adopt a baseline method that provides an absolute comparison across methods. The baseline has access to $S^{tr}$ and imputes $\tau$ for any element of $S^{te}$. To allow meaningful comparison to other methods, it employs the same imputation for missing scores, $S^m = \tau$.

*A note on LM:* Although the output of LM can be directly used for matching, it does not exploit observed suitabilities in its usual formulation. However LM can make use of some of the training data $S^{tr}$ by incorporating submitted papers assessed as "suitable" by some reviewer $r$ into her word vector $w_r$. Specifically, we include all papers in $w_r$ for which $r$ offered a score of 3 (only if this score is in $S^{tr}$).

For all methods, once an optimal match $x^*$ is found, we evaluate it using all observed and unobserved scores, with the same constant imputation for the missing scores, where *match quality* is measured using $J^{basic}$ (see Equation 3):

$$\sum_r \sum_p x^*_{rp}(S^{tr} \cup S^{te} \cup S^m = 1) \qquad (9)$$

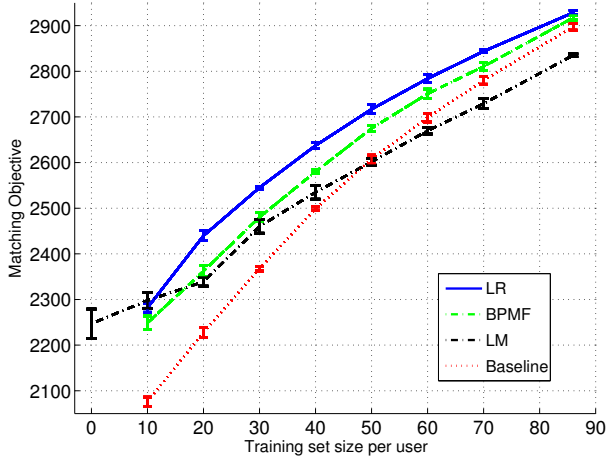

Figure 3: Performance on the matching task on the *N10* data set.

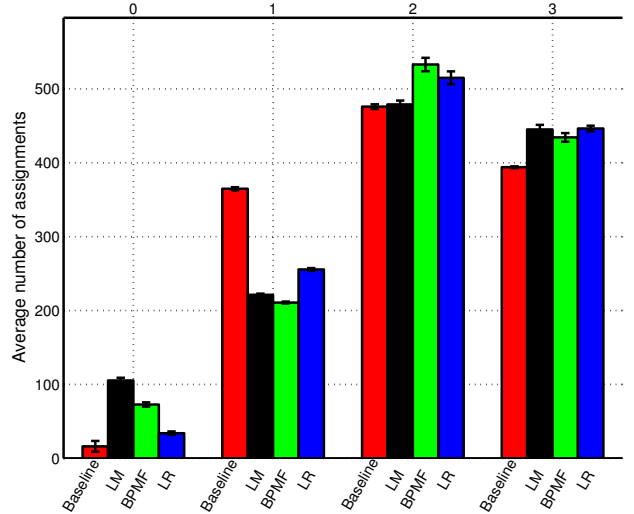

Figure 4: Assignments for *N10* by score value when using 40 training examples per user.

**Matching Performance using *Basic IP*:**  We now report on the quality of the matchings that result from using the predictions of the different methods. Again we consider dynamic matching performance as the amount of training data per user increases. Note that the optimal match value is 3053 for *N10* and 2172 for *N09*, which occurs when $\hat{S}^{te} = S^{te}$.

Figure 3 shows how matching quality varies as the amount of training data per user increases. Since training scores are also observed at matching time (Eq. 9), all methods benefit from having a larger training set. Figure 3 leads to the following three observations. Firstly, when no observed data is available, i.e., using only the archive, LM does very well, with a matching score of $2247 \pm 32$, nearly identical to the quality of LR and BPMF with 10 bids per user, and much better than the match quality of 1262 obtained using constant scores $((S^{te} \cup S^m) = \tau)$. Secondly, when very few scores are available, LR and LM perform best (and do equally well). As mentioned above, LM is able to exploit observed suitabilities by adding relevant papers to the user corpus, but this attenuates the impact of elicited scores: we see LM is outperformed by all other methods when sufficient data is available. Thirdly, LR outperforms all other methods as data is added. We also see that as the number of observed scores increases, unsurprisingly, the gain in matching performance (value of information) from additional scores decreases.

It is also interesting to note that a total matching score of over 2500 implies that, on average, each reviewer is assigned papers on which her average preference is greater than 2 (out of 3). LR reaches this level of performance with less than 30 observed scores per user, while other methods need 30% more data per user to reach the same level of performance.

Further insight into matching quality on *N10* induced by the different learning methods can be gained by examining the distribution of scores associated with matched papers (Figure 4) or under different sizes of the training set (Figure 5). Figure 4 displays the number of scores of each value (0–3) that get assigned with a training set size of 40. Not surprisingly, LR and BPMF assign significantly more 2s and 3s combined than all other methods. LM is very good at picking the top scores which reinforces the fact that word-level features, from reviewer and submitted papers, contain useful information for matching reviewers. Similar results were obtained on *N09* and thus LM's performance is not only a consequence of the data collection method used for *N10*. In addition, Baseline assigns few zeros, since all missing and test scores are imputed to be $\tau = 1$.

Figure 5 provides another perspective on assignment quality. Here we plot results for the best performing method, LR, on both *N10* and *N09*, for 3 different training set sizes. We first note that the extreme imbalance in *N09* leads LR to assign many zeros even with 80 training scores per user. Overall, both data sets show that as the number of training scores increases, more 2s and 3s, and fewer 0s and 1s, are assigned.

Our remaining results deal exclusively with *N10* since experimental results with *N09* were similar.

**Load Balancing *Balance IP*:**  The experiments above all constrain the number of papers per reviewer to be within a specific range ($P_{min}$–$P_{max}$). There is no good indication as to how to set these two extrema. Instead we now use the *Balance IP*, both for matching and evaluation (see Eq. 9), setting $f$ to be the absolute value function.

Figure 6 shows the histogram of assigned papers per re-

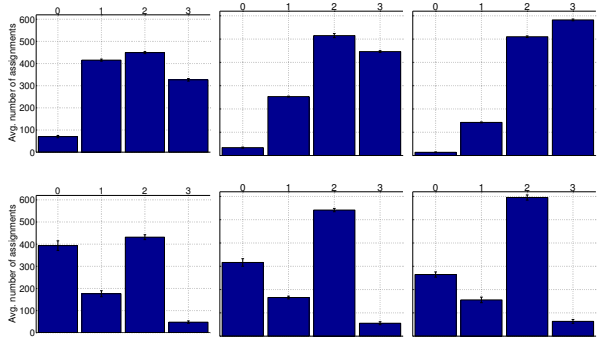

Figure 5: Assignments by score value. From Left to Right: 10, 40, 80 (86 examples for *N10*) training examples per user. Top: *N10*. Bottom: *N09*.

viewer given by the optimal solution to the IP for different $\lambda \in \{0, 0.1, 1\}$. When $\lambda = 0$ load equity is ignored, and almost all reviewers either get assigned the minimum ($R_{\min}$) or the maximum ($R_{\max}$) number of papers; within-reviewer variance ($\sum_p (x_{rp} - \bar{x})^2 / M$) is extremely high. When a "soft constraint" on load equity is introduced, assignments become more balanced as the $\lambda$ increases (i.e., the balance constraint becomes "harder"). The following table reports the matching objective versus the variance, averaged across users, for different values of $\lambda$ with a training set size of 40 (other training sizes yielded similar results):

| $\lambda$ | 0 | 0.1 | 0.25 | 0.5 | 0.75 | 1 |
|---|---|---|---|---|---|---|
| $J^{basic}$ | 2625 | 2615 | 2600 | 2573 | 2569 | 2569 |
| Variance | 4.62 | 3.28 | 2.61 | 0.89 | 0.37 | 0.33 |

Not surprisingly, larger penalties $\lambda$ for deviating from the mean reviewer load give rise to greater load balance (lower load variance) and worse matching performance. Generally, an appropriate $\lambda$ will be chosen by the conference organizers, that nicely trades off performance versus load balance across reviewers (here, perhaps around $\lambda = 0.5$).

### 4.4 Transformed Matching and Learning

We now consider a non-linear transformation of the scores, reflecting the view that it is *much* better to assign reviewer-paper pairs with suitabilities of 2 and 3, than pairs with 0 and 1; as discussed above this can be accomplished by allowing "utility" $x_{rp}$ to be non-linear in suitability score $s_{rp}$. We adopt the following sigmoid function to effect this non-linear transformation: $\sigma(s) = 1/(1 + \exp(-(s - 1.5)\beta))$; here 1.5 is the middle of the scores' range. We set $\beta = 4.5$, which gives: $\sigma(0) = 0.001$; $\sigma(1) = 0.095$; $\sigma(2) = 0.90$; $\sigma(3) = 1.0$. We first show how this transformation impacts matching performance without learning; then we discuss how one can incorporate the transformation into the learning objective itself.

We first test how matching using the transformed objectives affects results without using learning to infer missing scores (consequently, $S^u = \tau$), by examining difference in matching performance when varying the percentage of observed scores. Figure 7(a) shows the difference when matching with the transformed objective ($J^{tfm}$) versus the basic objective ($J^{basic}$). In both cases the resulting matches are evaluated using $J^{tfm}$. Although a minor gain is observed when most of the known data is observed, there is, overall, very little difference in performance when matching with either objective. Recall that the mean number of scores per paper is less than 4. Hence, when matching using a small fraction of the data, the matching procedure has very little flexibility to assign high scoring pairs unless learning is used to predict unobserved scores.

We can modify the learning objective to take into account the nonlinearity introduced in the matching objective. We do this by transforming all labels using the same sigmoidal transformation as in the matching objective (Eq.8). This allows learning to better predict the transformed scores by explicitly training on them Figure 7(b) shows the transformed matching performance of both LR on the non-transformed data, and LR-TFM, a linear regression model trained using the transformed learning objective. Not surprisingly, LR-TFM outperforms LR across all training set sizes, since it is trained for the modified objective $J^{tfm}$. The difference is especially pronounced with smaller training sets—when enough data is available, both methods will naturally assign many 2s and 3s. (We also verified that LR-TFM outperforms BPMF trained on the transformed objective).

## 5 Conclusion

We have developed a framework for paper-to-reviewer assignment in the context of scientific conferences. We showed how by eliciting only a small subset of scores from reviewers and inferring unobserved scores, using one of several learning methods, we are able to determine high quality matchings. Interestingly, in the field of collaborative filtering, side-information is often perceived to be useful only in the cold-start condition, where few or no scores are available. The performance of both LM and LR, which leverage word-level features from reviewers and submitted papers, show that this is not the case in our domain. We also explored the trade-off between matching quality and paper load balancing, which helps one avoid the need to manually set limits on the reviewer load. Finally we showed that using the realistic assumption that utility is non-linear in suitability score, we discover better matches using the same nonlinear transformation in the learning objective.

Given how matching benefits from an interaction with learning, we are developing ways to strengthen this interaction by making the learning methods sensitive to the final matching objective. We have obtained good results using this approach in an active learning setting where the system chooses which reviewer scores to query. We are also interested in exploring how asking meta-queries, about general aspects of papers rather than a single paper, may be exploited to reduce the number of reviewer queries while maintaining strong matching performance.

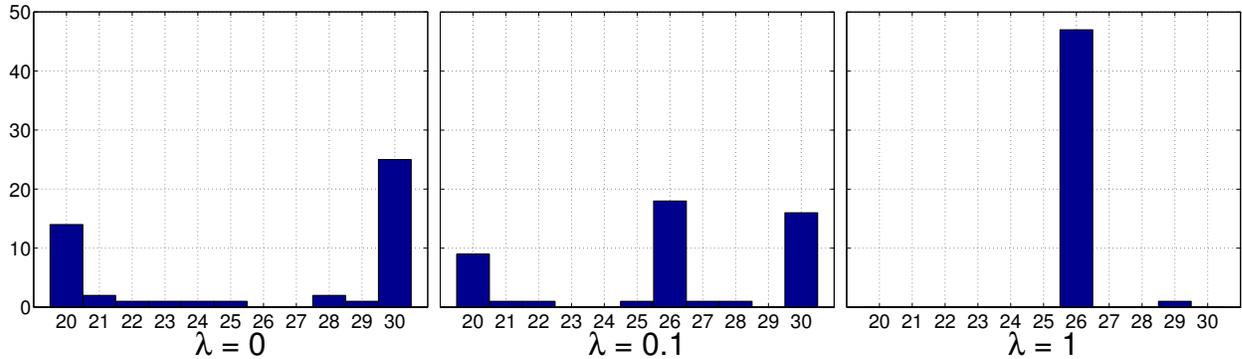

Figure 6: Histograms of number of papers matched per reviewer with different values of λ. The leftmost plot shows results with only hard constraints on reviewer loads ($\lambda = 0$); the others also include a soft constraint minimizing load variance.

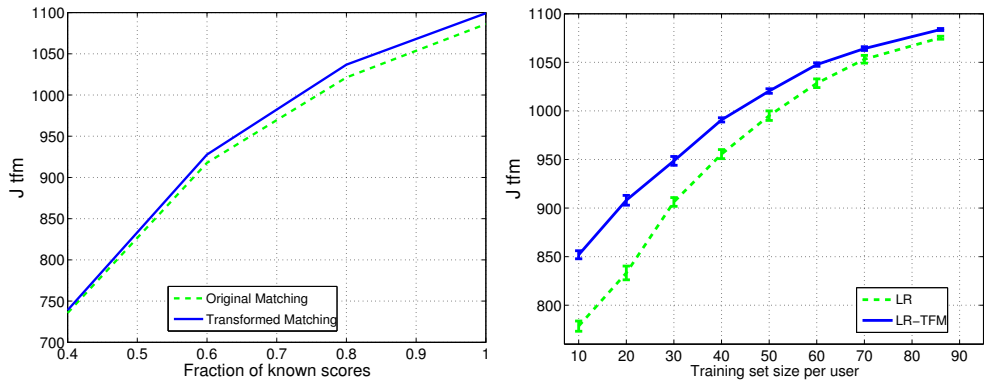

Figure 7: Performance on the transformed matching objective on *N10*. (a) Comparing performance of original ($J^{basic}$) and transformed ($J^{tfm}$) matching objectives, without learning. (b) Comparing performance of original and transformed LR learning using the transformed matching objective.

**Acknowledgements:** Special thanks to the NIPS Foundation and to the NIPS 2009 program chairs, Chris Williams and John Lafferty for their assistance in collecting the *N09* data set. Thanks to the reviewers for helpful suggestions. This research was supported by NSERC and CIFAR.